\PassOptionsToPackage{pdfpagelabels=false}{hyperref} 
\documentclass[a4paper,fleqn,usenatbib]{mnras}
\usepackage{graphicx,amsmath,amssymb}
\usepackage{mathtools}
\newcommand{\be} {\begin{equation}}
\newcommand{\ee} {\end{equation}}
\newcommand{\bea}{\begin{eqnarray}}
\newcommand{\eea}{\end{eqnarray}}
\newcommand{\obs}{obs}
\def\aap{A\&A}
\def\apj{ApJ}
\def\aj{AJ}
\def\apjs{ApJ S}
\def\mnras{MNRAS}

\def\icarus{Icarus}

\begin{document}
\title[Planetary systems and three-planet resonances]{Is the Orbital
Distribution of Multi-Planet Systems Influenced by Pure 3-Planet Resonances?}
\author[M. Cerioni et al.]{M. Cerioni$^{1}$\thanks{E-mail: 
matias.cerioni@mi.unc.edu.ar}, C. Beaug\'e$^{1}$ and T. Gallardo$^{2}$
\\
$^{1}$Instituto de Astronom\'{\i}a Te\'orica y Experimental, Observatorio 
Astron\'omico, \\
Universidad Nacional de C\'ordoba, Laprida 854, C\'ordoba X5000GBR, Argentina\\
$^{2}$Instituto de F\'{i}sica, Facultad de Ciencias, Udelar, Igu\'{a} 4225, 
11400 Montevideo, Uruguay
}
\date{Accepted 2022 March 21. Received 2022 March 16; in original form 2022 January 1}
\pubyear{2022}
\label{firstpage}
\pagerange{541--550}
\maketitle

\begin{abstract}
We analyze the distribution of known multi-planet systems ($N \geq 3$) in the 
plane of mean-motion ratios, and compare it with the resonance web generated by 
two-planet mean-motion resonances (2P-MMR) and pure 3-planet commensurabilities 
(pure 3P-MMR). We find intriguing evidence of a statistically significant 
correlation between the observed distribution of compact low-mass systems and the
resonance structure, indicating a possible causal relation. While resonance
chains such as Kepler-60, Kepler-80 and TRAPPIST-1 are strong contributors, most
of the correlation appears to be caused by systems not identified as resonance
chains. Finally, we discuss their possible origin through planetary migration
during the last stages of the primordial disc and/or an eccentricity damping
process. 
\end{abstract}


\section{Introduction}

Ever since the detection of the first Hot Jupiter \citep{Mayor.Queloz.1995}, the 
idea of \textit{in-situ} formation has been rightfully questioned. Due to 
stellar radiation, gas discs do not have enough material to form giant planets in
the region near the star. Today we know that during the early stages of 
planetary formation, planetary orbits can be modified through different 
processes such as planet-planet scattering and disc-induced migration. The 
latter is the result of gravitational interactions with the protoplanetary disc, 
and can allow for wide-orbiting planets (formed in the outer and gas-richer 
regions of the disc) to spiral inwards closer to the star.

Given two planets, a sufficiently smooth migration can drive the system into a 
\textit{two planet mean-motion resonance} (2P-MMR), which occurs when the 
orbital periods are related by a ratio of two small integers. In terms of the 
Keplerian mean motions, 2P-MMRs occur when:
\be
k_i n_i + k_{(i+1)} n_{(i+1)} \simeq 0
\label{eq1}
\ee
where $k_i$ and $k_{(i+1)}$ are integers. In resonances, the mutual 
gravitational interaction is enhanced and may modify the structure of the phase 
space by providing sources of orbital stability or instability, depending on 
the characteristics of the system and initial conditions. Therefore, 
resonances play an important part in shaping the overall orbital distribution
of many exoplanetary systems. 

Because migration rates may be different for nearby planets, the primordial 
period-ratios may sweep a range of values during the early evolutionary 
stages. Although in the case of divergent migration (where planet 
separation increases) planets are unlikely to be trapped in resonance 
\citep[e.g.,][]{Henrard.Lemaitre.1983} regardless of damping or excitation of 
eccentricities, the case of convergent migration (where planet separation
decreases) is different. Convergent migration can lead to permanent capture into
resonance \citep{Goldreich.Tremaine.1980, Lee.Peale.2002}, and is believed to be
the cause of most observed MMRs, such as that of Neptune and Pluto
\citep{Malhotra.1991}, the satellites of Jupiter and Saturn, as well as several
large-mass exoplanetary systems \cite[e.g.,][]{Lee.Peale.2002, Beauge.etal.2003,
Ramos.etal.2017}. In fact, it is believed that resonance capture in stable
solutions of first-order MMRs is highly probable provided sufficiently slow
smooth differential migration and low initial eccentricities 
\citep[see][]{Henrard.Lemaitre.1983, Beauge.etal.2006, Batygin.2015}.

\begin{figure*}
\centering
\includegraphics*[width=2\columnwidth]{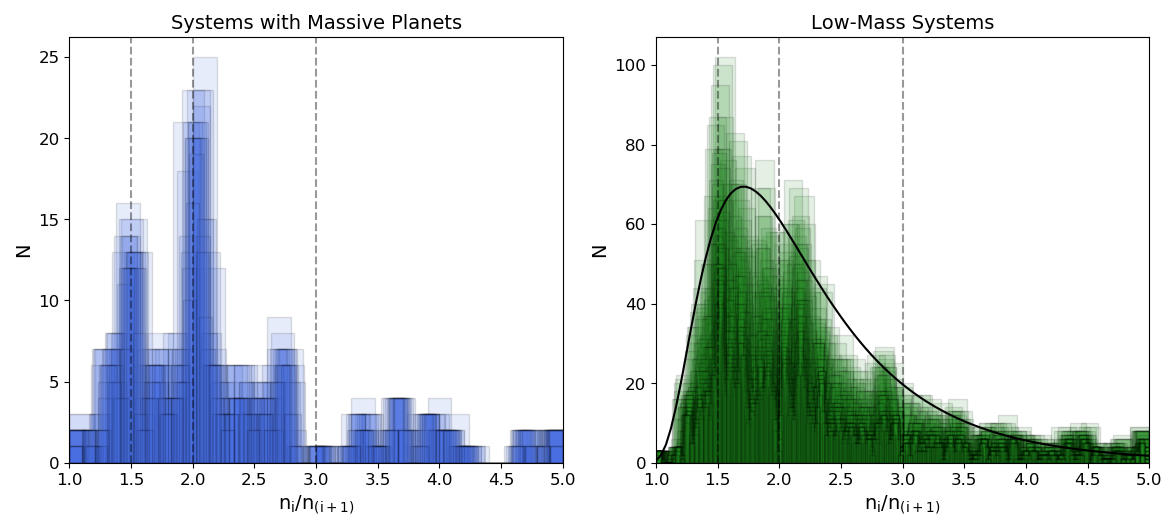}
\caption{Distribution of mean-motion ratios $n_i/n_{(i+1)}$ for adjacent 
planetary pairs belonging to systems with at least one giant planet (left) and
low-mass systems (right). The limit for the latter group was set to $20
m_\oplus$. Main MMRs are highlighted with dashed vertical lines. An analytical
lognormal approximation of the PDF for low-mass systems is shown as a black
curve. Its parameters are given in equation (\ref{eq5}).}
\label{fig_dist_n1n2}
\end{figure*}

If resonance capture were such a common outcome, or if such configurations 
were able to survive the dissipation of the protoplanetary disc, the 
distribution of mean-motion ratios $n_i/n_{(i+1)}$ between adjacent planets 
would show a strong correlation with the location of dominant commensurabilities 
(e.g. 2/1, 3/2). As it is well known, the real picture is more complex. Figure 
\ref{fig_dist_n1n2} shows the distribution of mean-motion ratios for 
two distinct populations of exoplanetary systems. The left-hand frame 
focuses only in those systems with at least one massive planet (defined as $m > 
100 m_\oplus$), while the data used for the right-hand plot only includes
low-mass systems where the largest known body has an estimated mass $m < 20 
m_\oplus$. As mass measurements are scarce among small planets, we opted
to employ the empirical mass-radius relation developed by
\cite{Chen.Kipping.2017}, and so these estimated masses will be prevalent in the
second category. Moreover, while both limits for the planetary mass may seem a
bit arbitrary, we found that the statistical analysis is fairly robust and does
not change significantly when adopting other values. Data was obtained from the
{\tt exoplanet.eu} catalogue and, as of July 2021, included a total of 830
multi-planet systems. Of these, only 104 contain at least one massive planet 
while most of the rest are part of the low-mass population.

In order to avoid possible problems constructing histograms with a small number 
of data points, we applied a smoothing technique, which consisted in overlaying 
a number of histograms with different number of bins $N_{\rm bins}$ and 
observing the resulting structure. Robust features would be repeated in many of 
the individual histograms and therefore appear plotted in dark tones. Spurious 
features, including spikes or gaps that only appeared for specific binnings, 
would be plotted only a few times and thus shown in lighter shades. We 
considered values of $N_{\rm bins}$ between $(N_{\rm data})^{1/2}$ 
and $4(N_{\rm data})^{1/2}$, where $N_{\rm data}$ is the total number of bodies 
in each group. These limits may again be considered fairly arbitrary, but we 
found no significant changes in the results for other choices.

The correlation between the observed $n_i/n_{(i+1)}$ and mean-motion resonances 
shows striking differences. Large-mass systems show a significant preference for 
values of orbital period ratios close to the 2/1 and 3/2 commensurabilities. In 
particular, the peak associated to the 2/1 resonance is very conspicuous. While
it is not possible to claim that most exoplanetary systems containing giant
planets are located in or close to MMRs, there is a clear abundance of resonant
systems. In contrast, we also find a partial gap in the vicinity of the
3/1 resonance.Previous works such as \cite{Fabrycky.etal.2014},
\cite{Delisle.etal.2014} and \cite{Bailey.etal.2022} provide further insight on
the impact of this MMR on the period ratio distribution.

The distribution of orbital period ratios in low-mass systems is, as expected,
more blurred with no clear preference for resonant over non-resonant
configurations. There appears to be a peak associated to the 3/2 resonance, but a
similar one for the 2/1 commensurability is not so striking. The plot shows that
most of the low-mass systems do not seem clustered around resonances.

\begin{figure*}
\centering
\includegraphics[width=2.1\columnwidth]{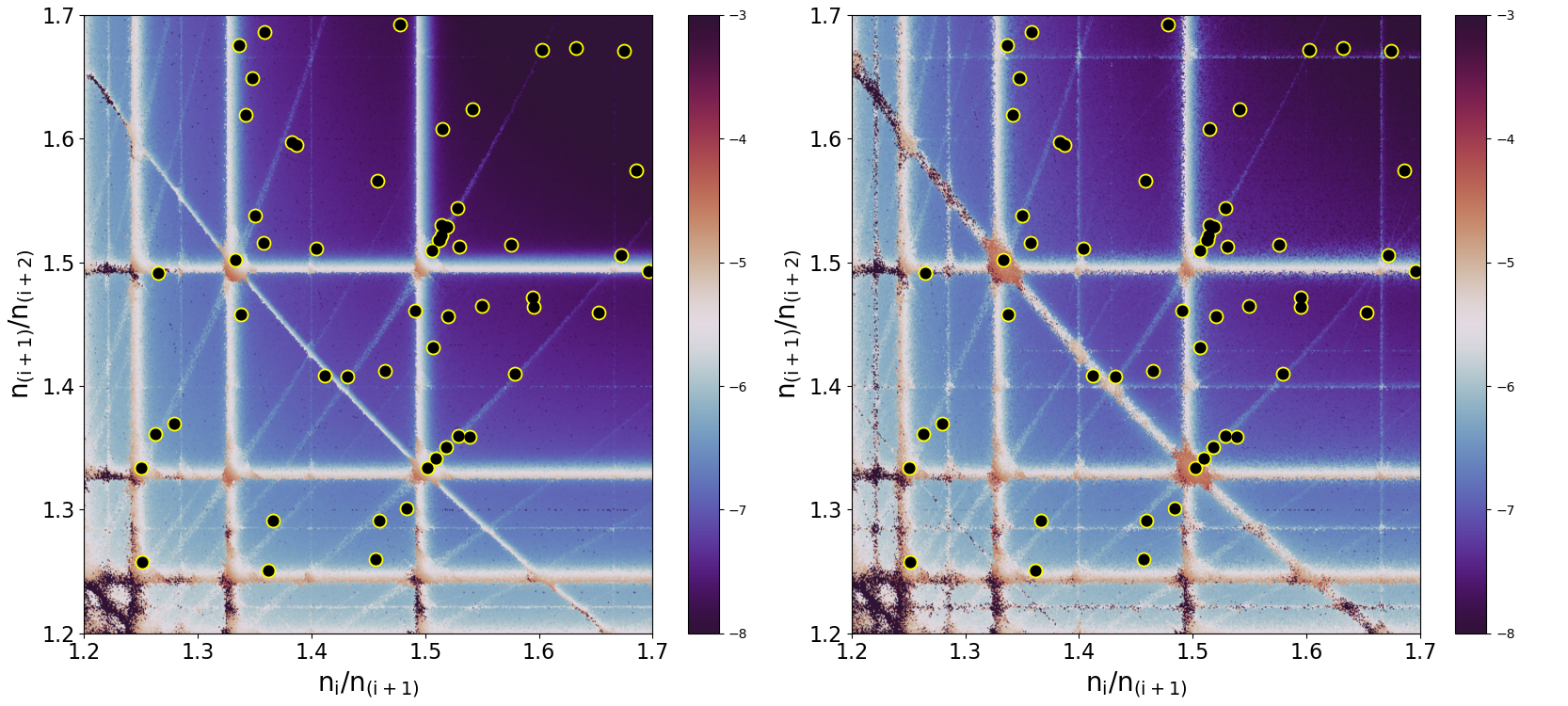}
\caption{Both plots show the distribution of mean-motion ratios for adjacent 
three-planet triplets belonging to low-mass systems. Data is overlaid to ${\rm
max}(\Delta a)$ dynamical maps, each constructed from N-body simulations of
$1000\times1000$ initial conditions, each integrated over a $T=10^4$ year
timespan. In both cases planetary masses for the maps were taken equal to
$m_1=m_2=m_3=10 m_\oplus$. The left-hand graph was constructed with initial
circular orbits, while the right-hand plot assumed $e_i=0.05$. All
initial angles were set equal to zero, except for $\lambda_i$ which were taken at
random for each different grid point.}
\label{fig_dist}
\end{figure*}

This result was initially surprising because disc migration theory
predicted that planet pairs would often be caught into resonances 
\citep{Goldreich.Tremaine.1980, Lee.Peale.2002}, although today we have a
few hypotheses that can explain the observed difference. One of them was
proposed by \cite{Goldreich.Schlichting.2014}, where they have shown that
eccentricity damping due to planet-disc interactions in addition to
planetary migration introduces an equilibrium eccentricity which may be
higher than the threshold allowed for resonant capture. This threshold is
dependent on planetary mass and as such, smaller planet would be less
likely to be captured. More recently,
\citep{Izidoro.etal.2017,Izidoro.etal.2021} found that many of the resonant
chains formed among low-mass planets could become unstable after the dissipation
of the disc, ejecting the systems from the commensurabilities and contributing to the observed low number of resonant configurations.

Notwithstanding these predictions, the last decade has seen the discovery
of a number of systems locked in multi-planet MMRs. While some cases have
also been detected among large-mass systems, most noticeably GJ876 and
HR8799, these dynamically complex configurations appear predominantly in
low-mass compact systems (e.g. Kepler-60, Kepler-80, Kepler-223,
TRAPPPIST-1, TOI-178), usually close to the central star. As of today
these multi-resonant systems appear as resonance chains whose building
blocks are either two-planet resonances (2P-MMR) or zero-order
three-planet resonances (3P-MMR). However, N-body simulations by 
\cite{Charalambous.etal.2018} and \cite{Petit.2021} have shown that capture in 
{\it pure} 3P-MMR configurations are also possible provided the migration rate 
is sufficiently low. We denote \textit{pure} 3P-MMRs as those where no
adjacent pair of planets lie in 2-planet commensurabilities.

\cite{Charalambous.etal.2018} in particular have shown cases where the 
end-result of migration of fictitious three-planet systems may include a 
combination of zero-order and/or first-order pure 3P-MMR which are distant from 
any two-planet commensurability. Such configurations would not be identified as 
resonant by studying the distribution of $n_i/n_{(i+1)}$ values and lead to the 
false impression that the system is in fact non-resonant.

Such diversity of stable multi-resonant configurations has motivated us to 
perform a more detailed analysis of the distribution of compact multi-planet 
systems, and search for any correlation with two-planet and three-planet 
commensurabilities. Our study is divided as follows. Section 2 discusses the 
known population of systems with $N \ge 3$ planets, their main dynamical 
features and distribution in the mean-motion ratio plane. The dynamically
relevant resonant population is described in Section 3, and the main statistical
analysis is shown in Section 4. Our main results are presented in Section 5 while
conclusions close the paper in Section 6.

\section{The Plane of Mean-Motion Ratios} \label{sec:MMRplane}

\subsection{Distribution of Known Systems} \label{sec:distro}

From the {\tt exoplanet.eu} catalog of confirmed exoplanets, we selected low-mass
systems with at least three planets ($N \geq 3$) and analyzed the distribution 
of mean-motion ratios $(n_i/n_{(i+1)},n_{(i+1)}/n_{(i+2)})$ for each 
sub-set of three adjacent planets. Thus, a 3-planet system would be 
characterized by a single data point in this plane, while a system with $N$
planets would yield $(N-2)$ distinct values. Since our study focuses on 
compact low-mass systems, we eliminated those triplets with coordinates larger 
than 1.7, thus removing the 2/1 MMR from the plot. The resulting distribution 
is shown in Figure \ref{fig_dist} and contains a total of 57 triplets, which we 
will denote by the set ${\cal P}_{\rm obs}$. 

The data points are overlaid on top of ${\rm max}(\Delta a)$ dynamical maps, 
each drawn from the numerical simulation of two different grids of $1000 \times 
1000$ fictitious 3-planet systems with orbits around a $1\ M_\odot$ star.
Each simulation was performed employing a Bulirsch-Stoer integration scheme with
variable time-step and precision $ll = 12$ \citep{Bulirsch1966}. Initial
conditions for the left-hand plot correspond to circular orbits, while in the
right-hand frame we opted for aligned orbits with $e_1=e_2=e_3 = 0.05$.
These are close to the mean value of known eccentricities for compact systems. 
In both cases the planetary masses were $m_1=m_2=m_3 = 10 m_\oplus$,
while the mean longitudes where chosen randomly between zero and $2
\pi$.  The ${\rm max}(\Delta a)$ structure indicator was chosen for its ability
to detect both 2P-MMRs and 3P-MMRs of any order. 

As shown in \cite{Charalambous.etal.2018}, each structure (line segment) in the 
dynamical maps is associated to a different mean-motion resonance. Vertical and 
horizontal strips correspond to two-planet commensurabilities, while 
intersection points between a horizontal and a vertical line mark the location 
of a three-planet resonance chain. Pure 3P-MMRs appear as diagonal curves 
stemming from bottom-left to upper-right, while an opposite trend mark 
2P-MMRs involving the inner and outer member of the triplet. Some of these 
resonances are evident even for $e_i=0$ (left-hand plot) while others are only 
dynamically relevant for eccentric orbits (right-hand frame). However, most of 
the resonance network appears fairly robust and common to both maps.

\begin{figure}
\centering
\includegraphics[width=\columnwidth]{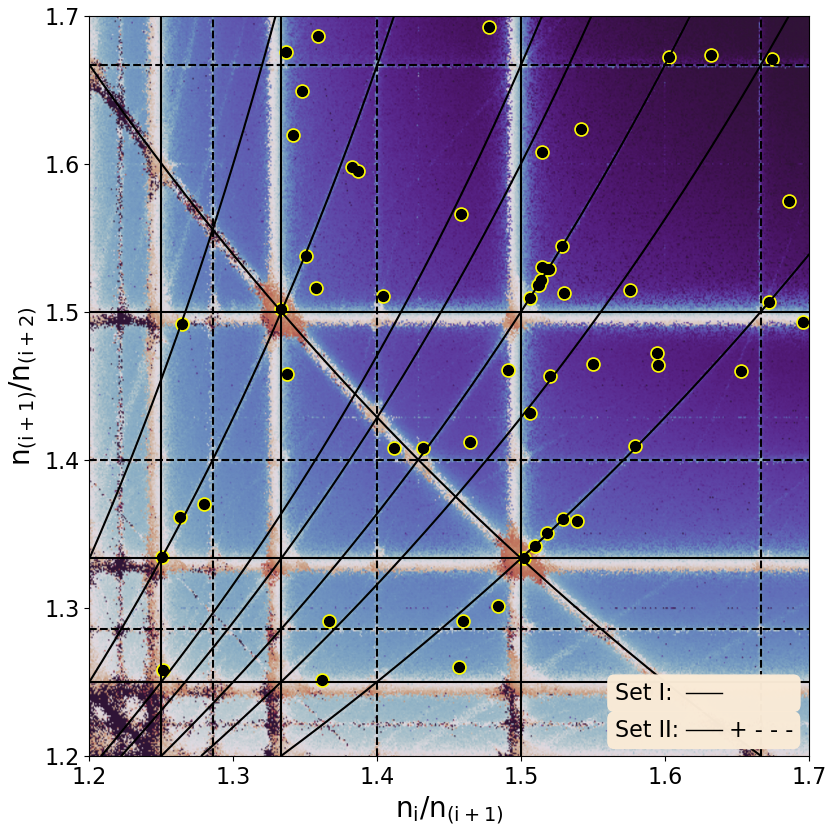}
\caption{Overlaid to the ${\rm max}(\Delta a)$ dynamical map and distribution 
${\cal P}_{\rm obs}$ of triplets, the plot shows two sets of MMRs used for 
the statistical analysis. {\bf Set I} is indicated by continuous black 
lines and includes first-order 2P-MMRs and zero-order 3P-MMRs. {\bf Set II} 
includes all of the above in addition to second-order 2P-MMRs. The complete list of resonances is summarized in 
Table \ref{tab1}.}
\label{fig_network}
\end{figure}

\subsection{The Resonance Network} \label{sec:network}

The two and three-planet resonances considered in this work are summarized in 
Table \ref{tab1} and drawn in Figure \ref{fig_network}. We defined two 
different networks. {\bf Set I} includes all first-order 2P-MMRs in the limits 
of the plane of mean-motion ratios (i.e. $n_i/n_{(i+1)}$ and 
$n_{(i+1)}/n_{(i+2)}$ between $1.2$ and $1.7$) as well as the most important 
zero-order 3P-MMRs. In addition, {\bf Set II} also includes 2nd-order 2P-MMRs in
the interval.The complete list is found in Table \ref{tab1}.

Three-planet resonances occur when the mean motions of a triplet of adjacent 
planets verify the relationship:
\be
k_{i} n_{i} + k_{(i+1)} n_{(i+1)} + k_{(i+2)} n_{(i+2)} \simeq 0
\label{eq2}
\ee
where $k_j$ are non-zero integers. The sum $s = k_{i}+k_{(i+1)}+k_{(i+2)}$ 
indicates the order of the resonance. Zero-order commensurabilities are 
typically the strongest \citep{Gallardo.etal.2016} and their libration domain is 
almost independent of the eccentricities \citep{Quillen.2011}. Through N-body
simulations, \cite{Charalambous.etal.2018} showed that zero and even first-order
3P-MMRs can guide the evolution of the system during migration, directing them 
towards a \textit{chain of resonances}, where the inner and outer pair would be 
in a 2P-MMR each. They found that for most planet-triplets, migration would halt 
once the chain was reached. Furthermore, \cite{Petit.2021} showed that a system 
can even be trapped in a first-order resonance, although this was only achieved 
for controlled conditions and robustness is uncertain.

The zero-order 3P-MMRs adopted for our work were chosen according to three
criteria. First, evidence of their dynamical effect must be noticeable in the 
${\rm max}(\Delta a)$ map, implying at least some significant orbital 
excitation within the integrated time-span. Second, the resonance must lie in a 
region of the plane inhabited by observed triplets. Although a rigorous 
definition will be given later on, for now it is sufficient to say that we will 
avoid the sectors close to the upper left-hand and lower right-hand corners. 
Finally, all the chosen 3P-MMRs must have the largest values of resonance 
strength in the region, as estimated with the semi-numerical method of 
\cite{Gallardo.etal.2016}. The 3P-MMRs listed in Table \ref{tab1} satisfy these
three conditions and all are of zero order.

The method for the calculation of the strength of a pure 3P-MMRs can be
summarized as follows. The Hamiltonian of the system $H$ can be written as the
sum of two parts $H = H_0 - R$; the first one, $H_0$, is the integrable part and
leads to the Keplerian motion of the planets around the central star, while $R$
groups all perturbations arising from mutual gravitational interactions between
the planets: $ R = R_{12} + R_{23} + R_{13}$, where 
\begin{equation}
R_{ij} = Gm_im_j \left(\frac{1}{|\mathbf{r_i-r_j}|}-\frac{\mathbf{r_i\cdot r_j}}{r_j^3}\right)
\end{equation}
denotes the mutual perturbations between planets $m_i$ and $m_j$. For a given
pure 3P-MMR, the critical angle is defined as
\begin{equation}\label{sigma}
\sigma = k_1\lambda_1 +  k_2\lambda_2 +  k_3\lambda_3 + \gamma
\end{equation}
where $\gamma$ is a combination of the fixed longitudes of the perihelia and
nodes of the three planets. Note that in the case of zero-order resonance
$\sigma$ is independent of $\gamma$. The resonant disturbing function,
$\mathcal{R}(\sigma)$, is constructed by an averaging method, fixing $\sigma$ and
calculating the mean of $R$ but including the effects of mutual planetary
perturbations. The disturbing function of a 3P-MMR is a second order function of
the planetary masses, which means the calculation of $\mathcal{R}(\sigma)$
cannot be done over $R$ evaluated at the unperturbed planetary positions. 

To properly calculate the mean it is necessary to take into account their
mutual perturbations in the position vectors. Accordingly, given any set of the
three planetary position vectors $\vec{r_i}$ satisfying the resonant condition
given by equation \ref{sigma}, the mutual perturbations of the three bodies are
computed and the $\Delta\vec{r_i}$ that they generate in a small interval $\Delta
t$ are calculated. These $\Delta\vec{r_i}$ produce a $\Delta R$, whose mean
$\mathcal{R}(\sigma)$ is then evaluated. This process involves considering a
large number of the three position vectors $\vec{r_i}$ which are always chosen
for a fixed value of $\sigma$. By changing this angle, $\mathcal{R}(\sigma)$ is
numerically constructed and finally the \textit{strength} is defined as the
semi-amplitude of $\mathcal{R}(\sigma)$. Additional details may be found in
\cite{Gallardo.etal.2016}.

\begin{table}
\caption{Two-planet and pure zero-order 3P-MMRs considered in the resonance 
networks {\bf Set I} and {\bf Set II}. Two-planet resonances are characterized by
the index array $(k_{i},k_{(i+1)})$, while 3-planet commensurabilities are
defined by the array $(k_{i},k_{(i+1)},k_{(i+2)})$. The (1,0,-2) resonance is a
2-planet commensurability involving the inner and outer member of the triplet. 
Both sets share the same 3P-MMRs. Additionally, we will also define {\bf
Set 0} as that containing the same 3-planet resonances but without any 2P-MMRs.}
\begin{center}
\begin{tabular}{cc|cc}
\hline
\multicolumn{2}{c}{{\bf Set I}} & \multicolumn{2}{c}{{\bf Set II}} \\
\hline
2P-MMR & 3P-MMR & 2P-MMR & 3P-MMR \\
\hline
  $(5,-4)$  &  $(1,-2,1)$  &  $(5,-4)$  &  $(1,-2,1)$  \\
  $(4,-3)$  &  $(1,-3,2)$  &  $(4,-3)$  &  $(1,-3,2)$  \\
  $(3,-2)$  &  $(2,-5,3)$  &  $(3,-2)$  &  $(2,-5,3)$  \\
$(1,0,-2)$  &  $(3,-7,4)$  &  $(9,-7)$  &  $(3,-7,4)$  \\
            &  $(4,-9,5)$  &  $(7,-5)$  &  $(4,-9,5)$  \\
            &  $(3,-8,5)$  &  $(5,-3)$  &  $(3,-8,5)$  \\
            &  $(5,-9,4)$  & $(1,0,-2)$ &  $(5,-9,4)$  \\
\hline
\label{tab1}
\end{tabular}
\end{center}
\end{table}

\begin{figure}
\centering
\includegraphics[width=\columnwidth]{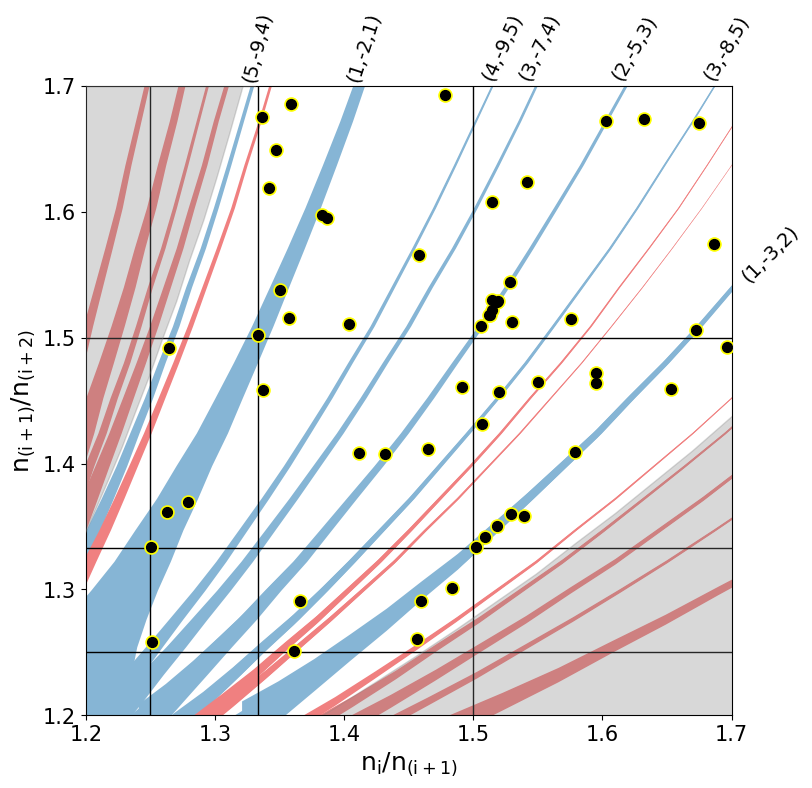}
\caption{Location of several pure 3P-MMR in the plane of mean-motion ratios. 
The thickness of each curve is not related to the width of the libration
domain but are proportional to the resonance strength, as estimated from
\protect\cite{Gallardo.etal.2016}. Commensurabilities chosen for our statistical study
are shown in blue. Black circles show the data set ${\cal P}_{\rm obs}$ of
triplets from observed planetary systems.}
\label{fig:fuerzas}
\end{figure}

Figure \ref{fig:fuerzas} plots the location of several zero-order 3P-MMRs,
whose thicknesses are drawn proportional to their strengths; the base
scale chosen to allow for a visual comparison. While the strength is a function
of the mutual distances, a subset of seven commensurabilities stand out over the 
other. These appear in blue and were chosen for our resonance network. The 
strongest ones appear to be those characterized by the index array 
$(1,-2,1)$ and $(1,-3,2)$ and house several triplets in their immediate 
vicinity. Other strong resonances, such as the $(1,-4,3)$ and $(3,-5,2)$ are 
located close to the edge of the plane and empty of known systems. Stability 
considerations probably favor the accumulation of data points in the 
general vicinity of the central area. In this region the $(2,-5,3)$ 
commensurability has a significant following, at least partially fueled by the 
$3/2$ double resonance.

\section{Correlation Between \texorpdfstring{${\cal P}_{\rm \obs}$}{} and the Resonance Sets}
\label{stat_analysis}

The main objective of this work is to establish whether the observed 
distribution of triplets ${\cal P}_{\rm obs}$ may have been determined or (at 
least) influenced by 2-planet and 3-planets resonances. In other words, 
we wish to study whether ${\cal P}_{\rm obs}$ is in some way correlated with 
the resonance network or if it is consistent with any random distribution. 
This approach is similar to that used to test the correlation of planetary 
pairs to 2P-MMRs; the main difference is that we must now analyze resonance
structures in a two-dimensional plane instead of a one-dimensional line segment.

In order to perform these tasks, we must first define two key aspects: (i) how 
to measure the correlation of the observed distribution to the resonance 
network, and (ii) how to check whether such a value is statistically 
significant. Each is discussed below.

\subsection{The Proximity Index}\label{sec:l_p}

In order to measure the correlation between a given sample and the resonance 
web, we will define a series of proximity indices as follows:
\be
I_p = \frac{1}{N}\sum\limits_{i=1}^N \Delta_i,
\label{eq3}
\ee
where $N$ is the total number of triplets ($N=57$ in our study) and 
$\Delta_i$ is the linear distance from the $i$-th triplet to its nearest 
resonance. If every point in a given sample were to fall exactly over a
resonance, then the calculation for that sample would yield $I_p = 0$.
Increasing values imply a smaller correlation between the observed distribution
and the resonance web.

Two aspects must be taken into consideration. First, while the location of each 
resonance in the plane of mean-motion ratios is defined in terms of mean 
elements, the values in ${\cal P}_{\rm obs}$ are osculating and thus their 
short-periodic variations have not been eliminated by any averaging process. 
While such a transformation may be carried out, even in the case of multi-planet
systems \citep[e.g.,][]{Charalambous.etal.2018}, it requires at least reasonable
estimations of the planetary masses, an asset not readily available in compact 
systems. Fortunately, the difference in mean-motion ratios between osculating 
and mean elements is not significant when studying low-mass planets (e.g. $m_i 
\sim 10 m_\oplus$), and we may avoid this step entirely. 

A second issue to keep in mind is the width of each resonance. While analytical 
models for both 2P-MMRs and zero-order 3P-MMRs \citep[e.g.,][]{Quillen.2011, 
Hadden.2019} are available, the libration widths are also function of the masses
as well as of the eccentricities. Since the values of $e_i$ are also poorly 
delimited, we preferred to disregard the libration width of all resonances and 
measure the distance from the position of each triplet to the centre of the 
nearest MMR. While this will introduce uncertainties in the calculations, we 
preferred a limited but (more) robust model than a more complex version 
strongly dependent on unreliable parameters.

\begin{figure*}
\centering
\includegraphics[width=2\columnwidth]{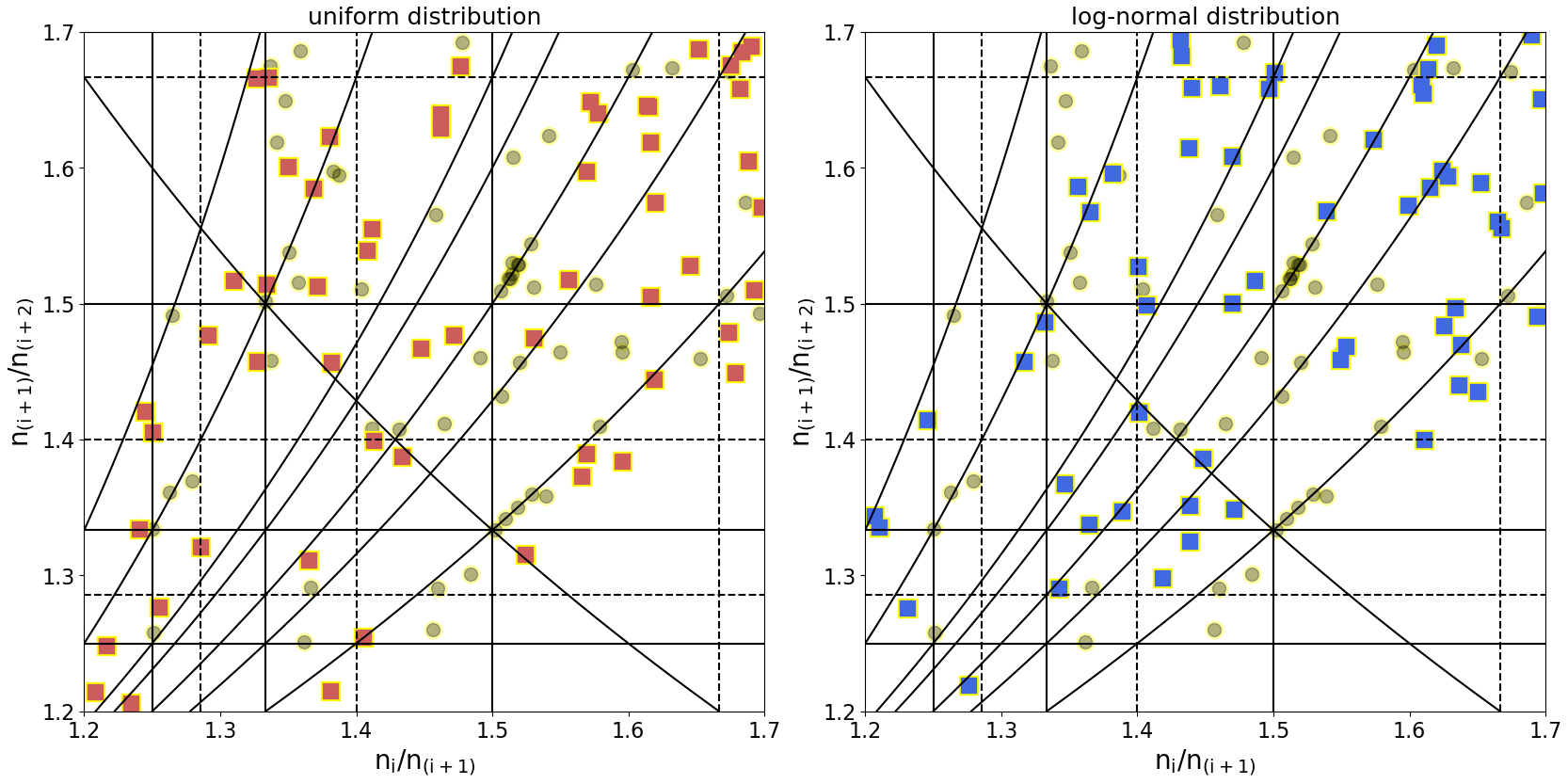}
\caption{Squares show examples of synthetic populations of triplets in the
mean-motion ratio plane. For comparison, the distribution of real planets are
shown in circles. The left-hand frame was drawn assuming an uniform distribution
of mean-motion ratios, while a log-normal PDF was considered in the right-hand
plot.}
\label{fig:ejsin}
\end{figure*}

\subsection{Random Synthetic Distributions \texorpdfstring{${\cal P}_{\rm rand}$}{}} 
\label{sec:distros}

Having calculated the values of the proximity index for the observed population,
i.e. $I^{(\rm obs)}_p$, we must now compare it with those determined from 
random distributions of triplets in the plane of mean-motion ratios. 
We will define each ${\cal P}_{\rm rand}$ as a set of $N=57$ data points in the 
plane of mean-motion ratios, such that each coordinate is drawn from a given
probability density function $f(x)$. We will adopt two different functional 
forms for the PDF. The first will be to assume a uniform distribution (i.e. flat
PDF) of $x$ within the range covered by our analysis (e.g. Figure 
\ref{fig_dist}). While simplistic in nature, it is also the most robust and 
requires no ad-hoc assumptions. We will also test a second PDF that reproduces 
the general trend of the observed exoplanetary population. Such a "background" 
PDF is estimated fitting a log-normal distribution to the mean-motion ratios of 
compact systems in the interval $x = n_i/n_{(i+1)} \in [1,4]$. This
range is wider than our main region of interest (1.2-1.7), and was chosen as
such in order to avoid over-fitting peaks and valleys associated with the main
2P-MMRs (namely the 2/1 and 3/2). Additionally, we tested different values for
the upper bound of the interval (ranging from 3 to 5), but found no significant
difference in the results. Thus, the fitting over our domain of interest seems
robust under small variations of the tail of the distribution, which reassures
us that, even though the falloff of planetary pairs with large period-ratios is
likely biased due to observational constraints, this should have little impact
on our results.

\textcolor{black}{
The (3-parameter) lognormal distribution is a continuous one-dimensional
distribution with the density function}
\be
f(x) = 
\begin{dcases}
\frac{1}{\sqrt{2\pi} \sigma (x-x_0)} \exp{\left[ -\frac{\left[ 
\ln{(x-x_0)} - \mu \right]^2}{2 \sigma^2} \right] }, \quad{x > x_0}\\
\\
0, \quad{x \le x_0} \\
\end{dcases}
\label{eq4}
\ee
where $\mu$, $\sigma$ and $x_0$ are the scale, shape and location
parameters, and they are such that if the random variable $X$ is lognormal, then
$Y = \ln(X - x_0)$ follows a normal distribution $Y \sim N(\mu,\sigma^2)$
\citep{Kozlov2019}. We adopted the values which maximized the likelihood
function (MLE), giving
\be
\mu = 0.159 \;\;\; \sigma = 0.485 \;\;\; x_0 = 0.804 ,
\label{eq5}
\ee
and the black curve in the right-hand frame of Figure \ref{fig_dist_n1n2} plots 
this function on top of the observed distribution. Except for localized peaks 
(associated with 2-planet MMRs) and slightly shallower valleys, the general 
agreement is very good.

\begin{figure*}
\centering
\includegraphics[width=2\columnwidth]{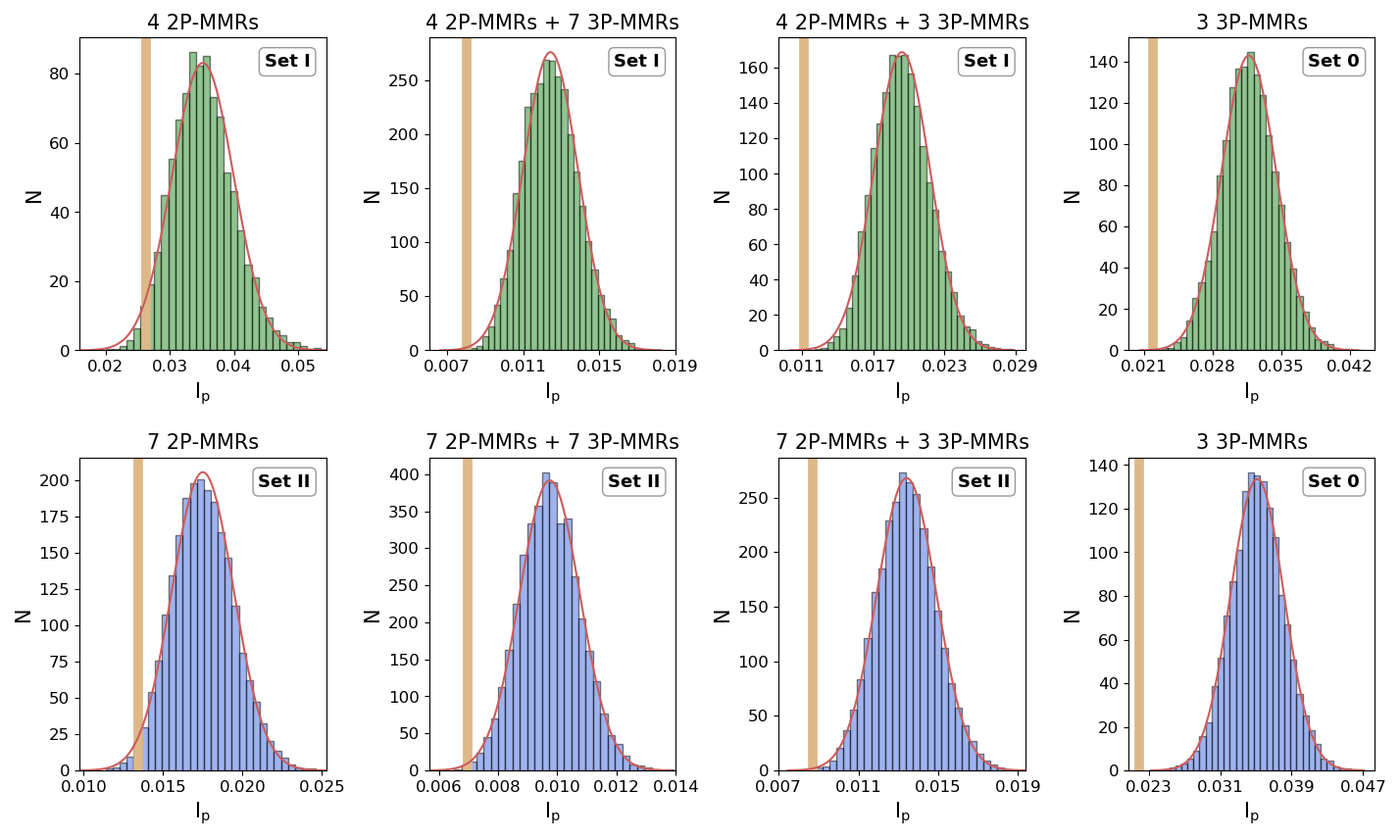}
\caption{Histograms show the distribution of proximity indexes $I_p$ for series 
of $10^4$ synthetic populations of $N=57$ random triplets in the mean-motion 
ratio plane. The set of resonances adopted for each graph is indicated in the 
upper right-hand corner, and the title shows how many resonances from
that set were used. Top plots assumed a uniform PDF for the random triplets,
while a log-normal distribution was adopted for the bottom graphs. The value of
$I_p$ obtained for the observed population (i.e. $I_p^{\rm (obs)}$) is
highlighted with a broad orange vertical line.}
\label{fig:Ip8}
\end{figure*}

Adopting such a smooth distribution as our null hypothesis is equivalent to 
assuming that the primordial formation sites of the planets were not influenced 
by mean-motion resonances, but primarily defined by global properties such as 
the surface density profile of solids and gas, disc evolution and short-term 
interaction between growing embryos. Thus, deviations from such a smooth and 
structure-less distribution, specially the overabundance of data points close 
to resonances, is considered a consequence of the post-formation dynamical 
evolution of the system. As mentioned in the introduction, the orbital
distribution of systems with large-mass planets shows significant peaks that
deviate from any attempt at a smooth fit and, consequently, a large number of
systems are found in the vicinity of 2P-MMRs. A similar correlation is more
difficult to establish for low-mass systems, thus our present study will not
only include 2-planet commensurabilities but also pure 3P-MMRs.

Both expressions adopted for the synthetic PDFs (uniform and log-normal) are 
one-dimensional; a random point in the plane of mean-motion ratios may then be 
obtained applying the distribution function to each coordinate. However, as
discussed previously and shown in Figure \ref{fig:fuerzas}, the observed 
population of exoplanets does not cover all the plane of mean-motion ratios, 
and seems to avoid regions where the triplets deviate significantly from 
equidistant configurations. This characteristic is reminiscent of the so-called 
peas-in-a-pod hypothesis and has been proposed several times over the past 
few years 
\citep[e.g.][]{Ciardi.etal.2013,Weiss.etal.2018b,Gilbert.Fabrycky.2020}. In 
order to incorporate this feature into our random synthetic populations, we will
only consider random data points if the values of 
$(n_i/n_{(i+1)},n_{(i+1)}/n_{(i+2)})$ lie within the white region of the 
plot. Curiously enough, the boundaries between the populated and empty regions
appear to roughly follow two different 3-planet resonances, the $(5,-9,4)$ and 
$(3,-10,7)$, and we will use both as a guide. Figure \ref{fig:ejsin} shows
examples of such synthetic populations (colored squares) overlaid to the
observed exoplanetary data (circles). 

\section{Monte-Carlo Simulations}
\label{sec:results-hist}

We generated two series of $10^4$ synthetic exoplanetary populations ${\cal 
P}_{\rm rand}$ in the plane of mean-motion ratios, each containing $N=57$ 
triplets; the first series was drawn from a flat PDF while the second used the 
log-normal distribution presented in the last section. For each random 
population we estimated the proximity index $I_p$, as defined by expression 
(\ref{eq3}), to each of the resonance groups {\bf Set I} and {\bf Set II}. See 
Table \ref{tab1} for their definitions. We added a third group (dubbed {\bf Set 
0}) that only contained the 7 pure 3P-MMRs but excluded all 2-planet 
commensurabilities. This set was used as a benchmark from which to evaluate how 
each type of resonance contributed to the global correlation index. 

The distribution of $I_P$ for each series of synthetic populations is shown as 
histograms in Figure \ref{fig:Ip8}, where the 
label and title in each frame indicate the resonance set and how many
resonances from it were used for the calculation respectively. Upper plots show
results obtained from a uniform PDF in the mean-motion ratio plane; analogous
runs adopting a log-normal distribution are presented in the bottom graphs. The
proximity indexes $I_p^{\rm (obs)}$ calculated for the observed distribution of 
exoplanetary systems are indicated with orange vertical lines.

In all cases the spread of synthetic $I_p$ may be well modeled by a normal 
distribution with mean $\mu$ and standard deviation $\sigma$, which we overlaid 
as a thin red curve over each histogram; no significant difference or deviation 
from the bell curve is observed. 

A great advantage of working with normal distributions is that it provides
straightforward tests to analyze the probability of obtaining a statistic 
$I_p \le I_p^{\rm (obs)}$ under the assumption that the null hypothesis is 
correct. For a normal distribution, this probability may be deduced by
performing a $Z$-test and calculating the standard score 
\be
Z = \frac{\mu - I_p^{\rm (obs)}}{\sigma} ,
\label{eq6}
\ee
which basically measures the distances of the observed index from the most 
probable value, in units of the standard deviation. 

Given $Z$, the probability of obtaining $|\mu - I_p| \ge |\mu - I_p^{\rm 
(obs)}|$ may then be estimated from the cumulative probability curve of the
normal distribution, and is given by
\be
P(|\mu - I_p| \ge |\mu - I_p^{\rm (obs)}|) = 1 - {\rm erf} \left( \frac{Z}{\sqrt{2}} \right),
\label{eq7}
\ee
where ${\rm erf}()$ is the error function. We can now use these statistics to 
analyze the results shown in Figure \ref{fig:Ip8}.

The two left-hand side plots compare the observed proximity index with random 
populations when considering only two-planet resonances. For {\bf Set I} and a 
uniform distribution of synthetic triplets the standard score is $Z = 1.84$, 
implying that the probability that the observed population of exoplanets 
satisfies the null hypothesis is $\simeq 0.07$. While relatively low, this
number is not sufficiently small to strongly suggest a correlation of the
planetary data with first-order 2P-MMRs. The lower left-hand plot shows similar
results, this time assuming a log-normal distribution of the random populations
and the {\bf Set II} resonance web. As before, only 2-planet resonances are
taken into account. The score of the observed distribution is now $Z = 2.1$ with
an associated probability of $P \simeq 0.04$. The exoplanetary population thus
shows little correlation with this extended resonance set and exhibits
practically the same average distance as any random distribution. This result is
consistent with other statistical studies performed over the past decade, and
the increase in confirmed systems does not appear to have changed the outcome. 

It is interesting to note that increasing the size of the resonance set does not
significantly affect the correlation with the observed distribution of triplets.
We believe this is additional evidence in favor of the null hypothesis and 
against any significant correlation with the 2P-MMRs web. Any sufficiently 
large set of lines in the mean-motion ratio plane would decrease the average 
distance to a set of points, independently of the orientation of the lines and 
of the data set. As more commensurabilities are considered, the distance between
any point and the web tends towards zero, and any dynamical structure that could
have been inferred from limited resonance webs would be lost. In this aspect,
our test proves robust and the results do not seem fooled by the number of
resonances being considered.

The second-to-left plots in Figure \ref{fig:Ip8} now take into account all the 
2-planet and 3-planet commensurabilities that make up {\bf Set I} (top) and 
{\bf Set II} (bottom). As before, for the top frame the random populations were 
drawn from a uniform distribution, while a log-normal PDF was assumed for the 
bottom graph. Both cases show significant changes generated by the 3P-MMR 
web and the standard scores are now $Z = 3.02$ and $Z = 2.72$, respectively. 
These yield probabilities of $P \simeq 0.002$ for the first set, and $P \simeq 
0.005$ for the second. Both values are very similar, indicating that any 
correlation is driven more by the web of 3-planet resonances and little 
affected by the 2P-MMRs and the underlying PDF for the random triplets. More 
importantly, results now indicate strong evidence in favor of a correlation 
between the observed distribution of planetary triplets and the resonance web 
with deviations from the norm of the order of $3 \sigma$. 

Next, we studied how the number of 3P-MMRs considered in the sets affects the 
correlation. The third column of Figure \ref{fig:Ip8} shows results when reducing
the number of 3P-MMRs to the three strongest resonances. As may be deduced from
Figure \ref{fig:fuerzas}, these correspond to the commensurability relations
\be
\begin{split}
\;\, n_i - 2 n_{(i+1)} + \;\, n_{(i+2)} &= 0 \\
\; n_i - 3 n_{(i+1)} +  2 n_{(i+2)} &= 0 \\
 2 n_i - 5 n_{(i+1)} +  3 n_{(i+2)} &= 0 . \\
\end{split}
\label{eq8}
\ee
Later on we will discuss how the results vary as function of the number of 
3-planet resonances. Results show an even higher degree of correlation between 
distribution and resonances, with now $Z = 3.48$ and $Z = 3.14$ for the upper 
and lower plots, respectively. The corresponding probabilities are $P \simeq 
0.0005$ and $P \simeq 0.0016$, or about an order of magnitude lower than with 
the complete sets. As with the 2P-MMRs, the degree of correlation increased with
a smaller resonance set, implying that most of them are probably not 
statistically relevant and do not contribute to the dynamical structure. 

\begin{figure}
\centering
\includegraphics[width=\columnwidth]{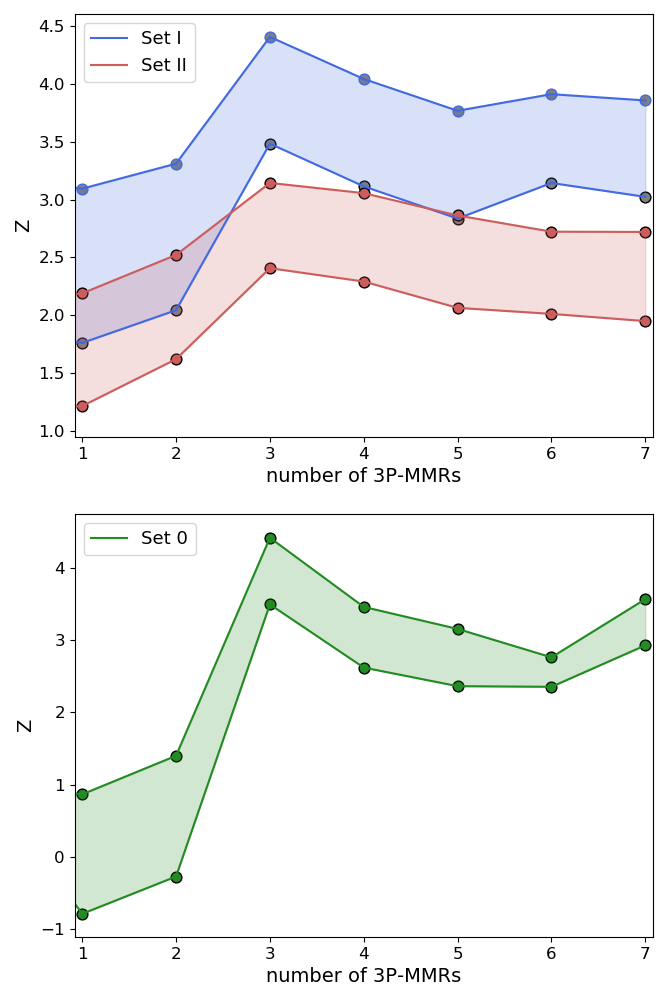}
\caption{Variation of the standard score $Z$ as a function of the number of 
3-planet resonances considered in the calculation. Upper and lower values 
correspond to synthetic populations following a log-normal or uniform PDF, 
respectively. The commensurabilities were ordered with decreasing 
strengths. For all resonance sets, only the first three 3P-MMRs appear 
relevant in establishing a correlation with the observed distribution of 
exoplanetary systems.} 
\label{fig:stat}
\end{figure}

As a final test, we considered only the 3-planet resonances in the plane, thus 
reducing the web to {\bf Set 0}. If the increase in number of 2P-MMRs from {\bf 
Set I} to {\bf Set II} decreased the correlation in all the previous 
calculations, we wondered whether 2-planet resonances were a relevant feature 
or a liability. Results are now shown in the right-hand top and bottom plots of 
Figure \ref{fig:Ip8}. The associated standard scores from the Z-test are $Z = 
3.50$ and $Z = 4.42$, while the probabilities fall to $P \simeq 0.0004$ and $P 
\simeq 0.00001$. Once again, the former were obtained assuming a uniform 
distribution of synthetic systems in the plane, and the latter (much lower) 
values were calculated from a log-normal PDF. 

Figure \ref{fig:stat} analyses how the standard score varies as a function of 
the number of 3P-MMRs considered in the statistical test. The resonances were
ordered according to their decreasing strengths, as estimated following 
\cite{Gallardo.etal.2016}. The top frame presents results for {\bf Set I} 
and {\bf Set II} while the bottom plot focuses on {\bf Set 0}. For each set, 
the top (bottom) values were obtained assuming a log-normal (uniform) 
distribution of triplets for each synthetic population. While we do not claim 
that the any other distribution will yield results somewhere between both 
limits, the shaded region helps visualize the general range of possible values.
As expected, all sets show similar trends, with a sharp increase in correlation 
up to the third strongest resonance, whose identities were revealed in equation 
(\ref{eq8}). Increasing the number of 3-planet resonances does not 
significantly alter the score, implying a more marginal effective correlation
with the observed distribution. However, we also do not find a noticeable
decrease in $Z$, so it is possible that some correlation does exist with the
larger set of resonances, even if not as striking. 

As a final caveat, it is important to keep in mind that we are working with 
low-number statistics. Even if we increase the number of synthetic populations 
and improve the modelization of the null hypothesis, the number of triplets per 
population is still small ($N=57$), a fact that undoubtedly affects the 
statistics. Spread over the mean-motion ratio plane, it is difficult to deduce what
would have been a better representative of the primordial averaged distribution,
and whether it would be better represented by a uniform or a log-normal
function. The difference in standard scores $Z$ obtained for {\bf Set 0} are
solely due to this issue and helps visualize the limitations of our analysis.
Perhaps all we can claim at this point is that the observed distribution of
planetary triplets appears significantly correlated with respect to the 3
strongest 3P-MMRs in the region. The probability that this correlation occurred
by chance lies somewhere between $0.001 - 0.1$ per cent. No significant correlation
with 2-planet resonances is observed.

\section{3P-MMRs vs Double Resonances}\label{sec:results-sigmas}

From a simple visual analysis of the distribution of known triplets (e.g. 
Figure \ref{fig:fuerzas}) it seems that a significant part of the correlation 
between the observed triplets and the resonance web could be caused by several 
multi-planet systems close to resonance chains (e.g. Kepler-80, TRAPPIST-1, 
etc), primarily located near coordinates $(3/2,4/3)$ and $(3/2,3/2)$. It is not 
clear whether these clusters effectively dominate the resonance correlation 
nor how other configurations contribute to the overall statistics. In this 
section we aim to analyze precisely this point and identify whether the 
correlation with 3P-MMRs is ubiquitous, or restricted to some subset of the 
observed population.

It is important to keep in mind that the limit between pure 3P-MMRs and double 
resonances is not always clear. The deviation of a given triplet from the 
intersection between two 2P-MMRs may be due to differential tidal effects 
acting over the age of the star 
\citep[e.g.][]{Papaloizou.2015,MacDonald.etal.2016}, or caused by the 
properties of the primordial disc that initially led to resonance capture 
\citep[e.g.][]{Charalambous.etal.2018}. The magnitude of the resonance offset 
\be
\Delta_{(p+q)/p} = \frac{n_i}{n_{(i+1)}} - \frac{(p+q)}{p}
\label{eq9}
\ee
depends on the driving mechanism as well as on the planetary masses, 
eccentricities and distance from the star. Offsets caused by tidal effects lead 
to values that increase over time and follow the position of 3P-MMRs. 
Conversely, no correlation with the 3-planet resonances is expected if 
the offsets are generated by strong eccentricity damping during planetary 
migration. 

Numerical simulations \citep[e.g.,][]{Ramos.etal.2017} usually lead to values up
to $\simeq 0.01$, at least for the 3/2 and 4/3 commensurabilities. This 
magnitude is consistent with known systems in double resonances, such as 
Kepler-60 \citep{Jontof-Hutter.etal.2016} with offsets 
$\Delta_{(p+q)/p} \simeq 10^{-4}$ and Kepler-223 \citep{Siegel.Fabrycky.2021} 
with $\Delta_{(p+q)/p} \simeq 10^{-3}$. Only for TRAPPIST-1 \citep{Teyssandier.etal.2022} and Kepler-60 
\citep{Jontof-Hutter.etal.2016} do 2-planet resonance offsets exceed $0.01$ and 
in some cases approach $0.02$. Many of the clusters observed in the 
mean-motion ratio plane extend far beyond these values, and it is not clear 
whether they actually correspond to double resonances or if they should be 
considered examples of true 3P-MMRs.

As a final note regarding this taxonomic issue, it is important to keep in 
mind that theoretical estimations of resonance offset have their limitations. 
Recent work on obliquity tides 
\citep[e.g.,][]{Millholland.Laughlin.2019,Millholland.Spalding.2020} 
show indications that planetary spins trapped in a type-2 Cassini state could 
accelerate the tidal heating and orbital in-fall, at least provided adequate 
precession rates from external perturbers. Moreover, current estimations of 
tidal timescales are usually based on classical constant time lag (CTL) models 
\citep[e.g.][]{Mignard.1979,Rodriguez.etal.2011}, which are probably not the 
most reliable for rocky planets. No analogous studies have yet been undertaken 
using more sophisticated rheological models 
\citep[e.g.,][]{FerrazMello.2013,Correia.etal.2014,Folonier.etal.2018}, which 
may also change our understanding of the problem and lead to larger values of 
$\Delta_{(p+q)/p}$ than currently predicted. 

\begin{figure}
\centering
\includegraphics[width=\columnwidth]{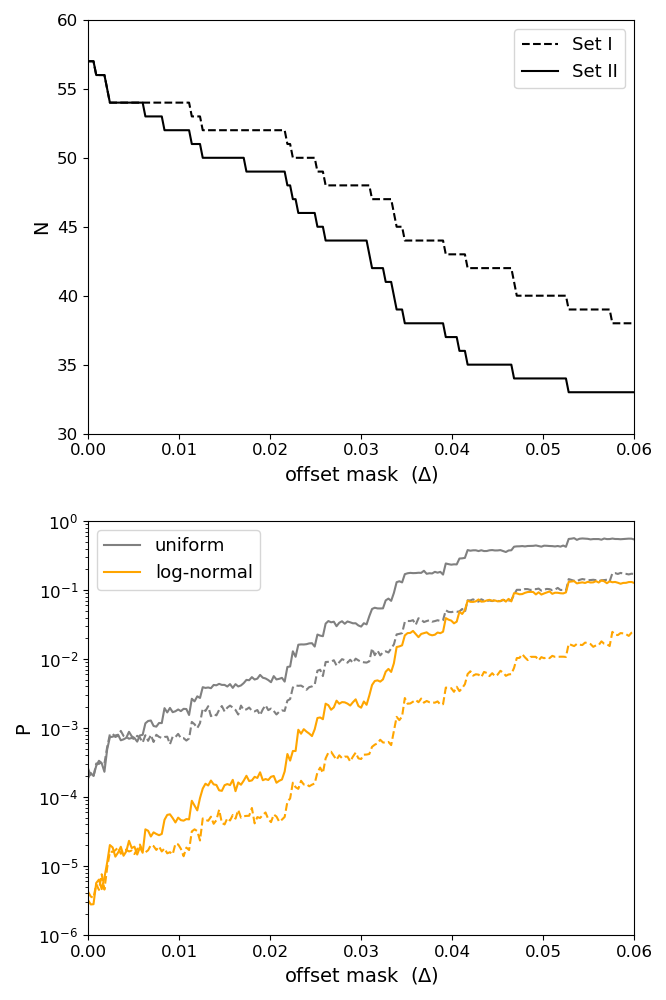}
\caption{Top frame gives the number of bodies of the observed population that 
have offsets larger than $\Delta$ for every possible double resonance in each 
set. The bottom plot shows the variation of the probability $P$ as 
function of the offset mask $\Delta$. Runs adopting the resonance {\bf Set I} 
are shown in dashed lines, while results for {\bf Set II} are indicated in 
continuous lines. } 
\label{fig:stat_sin}
\end{figure}

To analyze how (near) double resonance systems may affect our statistical 
analysis, we repeated our Monte-Carlo simulations using both {\bf Set I} and
{\bf Set II} but limiting the 3P-MMRs to the three strongest resonances, as
described in equation (\ref{eq8}). Because double resonances pertain to pairs of
adjacent planets, we also dropped the (1,0,-2) two-planet commensurability from
each set. We again generated several series of $10^4$ synthetic populations,
adopting both a uniform and a log-normal PDF, and compared the resulting
distribution of proximity indexes with that calculated with the observed
population ${\cal P_{\rm obs}}$. In each run, however, the observed population
was modified to exclude all triplets whose distance to the double resonances was
smaller than a predefined value $\Delta$. This {\it offset mask} was varied in
the range $\Delta \in [0,0.06]$, leading to a series of standard scores
$Z(\Delta)$ and probabilities $P(\Delta)$. 

Results are shown in Figure \ref{fig:stat_sin}, where the top plot shows how 
larger offset masks remove triplets and decrease the size of the observed 
population. The double resonances considered for each set are those defined by
every intersection of a 2P-MMR between $n_1$ and $n_2$ with another between
$n_2$ and $n_3$. For {\bf Set I} we considered 9 intersections including
$(3/2,3/2)$, $(3/2,4/3)$, $(5/4,3/2)$, etc. For {\bf Set II} the number of
double resonances increased to 49, causing a significantly faster decrease in
the number of surviving triplets as function of the offset mask.

The change in $P$ as a function of the offset mask is shown in the lower plot 
of Figure \ref{fig:stat_sin}, where now each color corresponds to a different 
PDF adopted for the random populations. Recall that $P$ is an estimation of the 
probability that the observed distribution of triplets stems from a random 
population uncorrelated with the resonance web. Low values are indicative of a 
high correlation, suggesting that the two and three-planet resonances have 
influenced the distribution of planetary triplets. Values of $P$ of the order of
unity point in the opposite direction. 

Independently of the resonance set, the observed distribution ${\cal P}_{\rm 
obs}(\Delta)$ shows a high correlation with the mean-motion commensurabilities, 
at least for offset masks up to $\sim 0.02$. This trend is more pronounced for 
a log-normal distribution of random populations, and lead to very low 
probabilities for offsets up to $\simeq 0.04$. In all cases all evidence of 
correlation ceases when the population decreases by $\sim 30$ per cent, 
which is expected given that the ability for a statistical test to
produce a low $P$ decreases with the sample size.

\section{Conclusions}

In this work we have presented a simple statistical analysis testing whether 
the distribution of known compact low-mass multi-planet systems shows
significant correlation with 2-planet and 3-planet resonances. As expected from
previous studies, only a faint correlation has been found with 2P-MMRs (Figure 
\ref{fig:Ip8}), and this result appears independent of the resonance set. 

Conversely, we do find intriguing evidence in favor of a correlation of the 
observed distribution with three-planet resonances, particularly a sub-set of 
three commensurabilities characterized by the index 
$(k_i,k_{(i+1)},k_{(i+2)})$ equal to $(1,-2,1)$, $(1,-3,2)$ and $(2,-5,3)$. 
These are the strongest pure 3P-MMRs in the region, a fact that increases the 
credibility of the result. If a set of pure 3-planet resonances were in some 
manner affecting the dynamics of exoplanetary systems, we would expect them to
be the strongest of the set. This is indeed the case. 

Among the systems which lie in 3P-MMRs, there are a striking number of
them which are also close to double resonances. A possible explanation could be
that these started as resonant chains which eventually proceeded to undergo
detectable divergence along pure 3P-MMRs due to tidal effects  \citep{Papaloizou.2015,Charalambous.etal.2018,Goldberg.Batygin.2021}.
Isolating these systems from the population reduces the correlation with
3P-MMRs, but does not erase the statistical significance, at least for offsets
of the order of those expected from tidal evolution and/or resonance capture
scenarios. This seems to indicate that although resonance chains such as
Kepler-60, Kepler-80, Kepler-223 and TRAPPIST-1 are an important factor, they
are not dominant and the observed correlation is also fueled by systems not
identified as resonance chains. Still, many of the systems showing strong
correlation to pure 3P-MMRs lie relatively close to double resonances, but with
large offsets of the order of $\Delta_{(p+q)/p} \sim 0.05$. 

Both \cite{Charalambous.etal.2018} and \cite{Petit.2021} have shown that 
resonance capture in zero-order and first-order 3P-MMRs far from double
resonances is possible as long as the orbital decay timescale is sufficiently
long. It is thus possible that the observed correlation could have occurred
during the last stages of the primordial disc when its surface density was very
low and orbital excursions very limited. A similar outcome would be obtained
assuming that the disc-planet interactions primarily affected the eccentricities
\citep{MacDonald.Dawson.2018}. In both cases the triplets would not have had 
sufficient time to evolve towards their final resting place (i.e. double 
resonances), leading to a distribution of systems with an excess of triplets 
close to but not exactly in resonance chains. Our results appear to be in favor 
of such a hypothetical scenario.

Finally, it is important to stress that the observed population is still small 
and we are working with low-number statistics. While the results are intriguing 
and could lead to a relevant link between pure 3-planet resonances and compact 
planetary systems, a much larger population is required to confirm this trend.

\section*{Data Availability}

The data underlying this article are available in Github, at
\href{https://github.com/matiaspcerioni/Cerioni-Beauge-Gallardo-2022}{https://github.com/matiaspcerioni/Cerioni-Beauge-Gallardo-2022}. 
The original data was obtained from {\it The Exoplanet Encyclopaedia}. 
The selected sub-samples as well as the scripts used to carry out the analysis
are available in these repositories.

\section*{Acknowledgements}
The authors would like to express their gratitude to the computing facilities of Instituto de Astronomía Teórica y Experimental (IATE) and Universidad Nacional de Córdoba (UNC), without which these numerical experiments would not have been possible. 
We would also like to express our gratitude to an anonymous reviewer for insightful suggestions that helped improve the work. 
This research was funded by research grants from Consejo Nacional de Investicaciones Científicas y Técnicas (CONICET) and Secretaría de Ciencia y Tecnología (SECYT/UNC). 
Planetary data were obtained from the portal {\tt exoplanet.eu} of The Extrasolar Planets Encyclopedia.

\bibliography{distribucion}{}
\bibliographystyle{mnras}



\end{document}